\documentclass{article}
\usepackage{spconf,amsmath,epsfig}
\usepackage{amssymb}
\usepackage{mathtools}
\usepackage{multirow}
\usepackage{array}
\usepackage{float}
\usepackage{dblfloatfix}
\usepackage{algorithm,algorithmic}
\usepackage{graphicx,subfigure}
\usepackage[dvipsnames]{xcolor}
\usepackage{ragged2e}
\usepackage{siunitx} 

\let\OLDthebibliography\thebibliography
\renewcommand\thebibliography[1]{
  \OLDthebibliography{#1}
  \setlength{\parskip}{0pt}
  \setlength{\itemsep}{0pt plus 0.3ex}
}

\begin{document}\sloppy

\def\x{{\mathbf x}}
\def\L{{\cal L}}

\newcommand{\todo}{\textcolor{red}} 
\newcommand{\yl}{\textcolor{blue}} 
\newcommand{\kb}[1]{\textcolor{purple}{#1}}

\title{alternating direction method of multipliers for negative binomial model with the weighted difference of anisotropic and isotropic total variation}

\name{%
   Yu Lu$^{\star}$%
   \qquad Kevin Bui$^{\dagger}$%
   \qquad Roummel F. Marcia$^{\star}$\thanks{This research is partially supported by NSF Grants IIS 1741490 and DMS 1840265.}%
}
\address{%
   $^{\star}$ Department of Applied Mathematics, University of California, Merced, California, USA \\%
   $^{\dagger}$ Department of Mathematics, University of California, Irvine, California, USA%
}

\maketitle

\begin{abstract}

In many applications such as medical imaging, the measurement data represent counts of photons hitting a detector. Such counts in low-photon settings are often modeled using a Poisson distribution. However, this model assumes that the mean and variance of the signal's noise distribution are equal. For overdispersed data where the variance is greater than the mean, the negative binomial distribution is a more appropriate statistical model. In this paper, we propose an optimization approach for recovering images corrupted by overdispersed Poisson noise. In particular, we incorporate a weighted anisotropic–isotropic total variation regularizer, which avoids staircasing artifacts that are introduced by a regular total variation penalty. We use an alternating direction method of multipliers, where each subproblem has a closed-form solution. Numerical experiments demonstrate the effectiveness of our proposed approach, especially in very photon-limited settings.

\end{abstract}

\begin{keywords}
Negative binomial distribution, Poisson distribution, alternating minimization, total variation
\end{keywords}

\section{Introduction}
\label{sec:intro}
The Poisson distribution has been widely seen in image processing, including medical imaging, night vision, and astronomy, but it strongly assumes equal mean and variance. In practice, although the noise distribution may appear Poisson, its mean and variance are typically different. For the case with overdispersed Poisson distribution where the variance is larger than the mean, we consider the negative binomial (NB) distribution as a more suitable model than the Poisson distribution. The probability mass function (p.m.f.) of the $\text{NB}(r,p)$ distribution is given by
\begin{align*}
     \mathbb{P}(y; r, p) =  \binom{r+y-1}{y}(1-p)^y p^r,
\end{align*}
where $y \geq 0$ is the number of successful events within a sequence of independent and identically distributed Bernoulli trials occurring before the $r^{\text{th}}$ failure event and $p \in [0,1]$ is the failure probability of each trial.
The mean of the $\text{NB}(r,p)$ distribution is given by $\mu = r(1-p)/p$, which implies $p = r/(r+\mu)$.  
It can be shown  that if $r \to \infty$ and $p \to 1$ in such a way that $\mu$ remains constant, the NB p.m.f.\ will tend to the Poisson p.m.f.\ (see \cite{degroot2012probability}). In  other words, the Poisson distribution is a special case of the NB distribution. 

Under the negative binomial assumption, our model for the components of the observation vector $g \in \mathbb{R}_+^m$ can be written as 
\begin{align*}
    g_i \sim 
    \hbox{NB}(r, p) = 
    \hbox{NB}\left (r, \frac{r}{r+(Af^{\star})_{i}} \right ), 
\end{align*}
where  $f^{\star} \in \mathbb{R}^{ n}_+$ is the signal or image of interest and $A \in \mathbb{R}_+^{m \times n}$ linearly projects $f^{\star}$ onto a set of expected measurements $Af^{\star} \in \mathbb{R}_+^m$.
Overall, the probability of observing $g$ given  the  linear projection $Af^{\star}$ is
\begin{align}\label{eq:pop}
     \mathbb{P}( g | Af^{\star} ) 
     \!=\!  \prod_{i=1}^{m} \!
     \binom{ r\!+\!g_i\!-\!1  }{g_i} \!\! 
     \left ( \! \frac{r}{r\!+\!(Af^{\star})_i} \! \right )^{\!\! r}
     \!\!\!  
     \left ( \! \frac{(Af^{\star})_i}{r\!+\!(Af^{\star})_i} \! \right )^{\!\! g_i}\!\!\!,
\end{align}
The corresponding negative log-likelihood function is given by 
\begin{align}
    F(f) \! \equiv \! \sum_{i=1}^m (r\!+\!g_i)\! \log(r\!+\!(Af)_i) \!-\! g_i\! \log((Af)_i).
\end{align}
(see \cite{hilbe2011negative} for details).

In image processing, employing regularization is a common technique for enhancing image recovery. Therefore, when recovering an image corrupted by NB noise, the general problem we aim to solve is the following: 
\begin{align}\label{eq:obj}
    \hat{f} \ = \  &  \ \underset{f \in \mathbb{R}_+^n}{\text{arg min}}  \ \ F(f) + {\tau} \mathcal{R}(f).
\end{align}
where  $\tau > 0$ is a regularization parameter and $\mathcal{R}$ is a regularization on the image $f$. In this paper, we focus on the weighted difference of anisotropic and isotropic total variation (AITV) \cite{lou2015weighted} as the regularization of choice, and we propose an alternating direction method of multipliers (ADMM) algorithm \cite{boyd2011distributed} to solve \eqref{eq:obj}.

\section{Related Work}

\subsection{Negative binomial model}
The NB distribution was established to be connected to the Poisson distribution as a useful model for overdispersed Poisson noise, where the variance is greater than the mean \cite{anscombe1948transformation,fisher1941negative, gardner1995regression}. It has been utilized in various applications, such as traffic accident analysis \cite{poch1996negative}, data regression \cite{allison20027}, and matrix factorization \cite{gouvert2020negative}. The NB distribution has been extended to signal and image processing \cite{lu2023negative}. Although the Poisson model has been used widely \cite{adhikari2015nonconvex, bui2023weighted, 
harmany2011spiral,   le2007variational}, the more general NB distribution promises to yield better results. 

\subsection{Regularization in image processing}
Total variation (TV) \cite{rudin1992nonlinear} has remained a classical choice for image regularization. It was employed successfully for recovering images under low-light conditions, e.g., images corrupted by Poisson noise \cite{bardsley2008efficient,le2007variational,willett2010poisson} or NB noise \cite{lu2023negative}. However, TV tend to create staircase artifacts or blur edges \cite{condat2017discrete}. 
These limitations prompted the exploration of various TV variants, such as TV$^p$ (where $0<p<1)$ \cite{chen2012non} and AITV \cite{lou2015weighted}. Among these variants, AITV outperformed classical TV and TV$^p$ for images corrupted by Gaussian noise \cite{lou2015weighted}. It also performed well in Poisson denoising \cite{bui2023weighted}. However, AITV has not yet been investigated for overdispersed Poisson noise or NB noise. 

\section{Proposed Model and Algorithm}
We aim to recover the image $f^*$ by solving the following optimization problem: 
\begin{align} \label{eq:proposed}
    \hat{f} \ = \  &  \ \underset{f \in \mathbb{R}_+^n}{\text{arg min}} \ \  F(f) + {\tau} \| f \|_{TV^{(\text{AI})}}, 
\end{align}
where AITV is defined as 
\begin{align*}
    \displaystyle \Vert f \Vert_{TV^{(\text{AI})}} = \Vert f \Vert_{TV^{(\text{A})}} - \alpha \Vert f \Vert_{TV^{(\text{I})}},
\end{align*}
where $\alpha \in [0,1].$
The anisotropic total variation (ATV) and the isotropic total variation (ITV) are given by
\begin{align*}
\displaystyle \Vert f \Vert_{TV^{(\text{A})}} &= \|\mathbf{D}f\|_1
= \sum_{i=1}^{n} \left( \vert (\mathbf{D}_x f)_{i} \vert+\vert  (\mathbf{D}_y f)_{i}\vert \right)
\end{align*}
and
\begin{align*}
\displaystyle \Vert f \Vert_{TV^{(\text{I})}}&= \|\mathbf{D}f\|_{2,1} = \sum_{i=1}^{n} \sqrt{ |(\mathbf{D}_x f)_{i}|^{2}+ |(\mathbf{D}_y f)_{i}|^{2}},
\end{align*}
respectively, where $\mathbf{D}_x$ and $\mathbf{D}_y$ are the horizontal and vertical difference operators on the
(vectorized) image $f$.  

To solve \eqref{eq:proposed}, we use an ADMM approach. First we introduce the auxiliary variables $v \in \mathbb{R}^{m}_+$ and $w \in \mathbb{R}^{n \times 2}$ such that $v = Af$ and $w= \mathbf{D}f = [\mathbf{D}_xf \ \ \mathbf{D}_yf]$. As a result, \eqref{eq:proposed} is reformulated as a constrained optimization problem:
\begin{gather}
\begin{aligned}\label{eq:problem}
    &\underset{f, v, w}{\text{minimize}} \ \ \sum_{i=1}^m 
    \bigg \{ 
    (r\!+\!g_i)\! \log(r\!+\!v_i) \!-\! g_i\! \log(v_i) 
    \bigg \} 
    \\
    & \hspace{3.65cm}  
    \!+\! \tau(\Vert w \Vert_{1} 
    \!-\!  \alpha \Vert w \Vert_{2,1} ) 
    \\ 
    & \hbox{subject to}  \ \   Af = v, \mathbf{D}f = w.
\end{aligned}
\end{gather}
The augmented Lagrangian of \eqref{eq:problem} is
\begin{align} \label{eq:lagrange}
    \mathcal{L}(f, v, w, x, z) = \ & \sum_{i=1}^m (r\!+\!g_i)\! \log(r\!+\!v_i) \!-\! g_i\! \log(v_i) \nonumber \\
    &+ \tau \left(\Vert w \Vert_{1} - \alpha \Vert w \Vert_{2,1} \right) \nonumber\\
    &+ \langle x, Af-v \rangle + \frac{\beta}{2} \Vert Af - v \Vert_{2}^2  \nonumber\\
    &+ \langle z, \mathbf{D} f - w \rangle + \frac{\beta}{2} \Vert \mathbf{D} f - w \Vert_{2}^2,
\end{align}
where $x \in \mathbb{R}^m$ and $z \in \mathbb{R}^{n \times 2}$ are the Lagrange multipliers and $\beta > 0$ is a penalty parameter. Therefore, the ADMM algorithm for \eqref{eq:proposed} iterates as follows:
\begin{subequations}
\begin{align}
    f^{k+1} \ = \  &  \ \underset{f \in \mathbb{R}_+^n}{\text{arg min}} \quad \mathcal{L}(f, v^k, w^k, x^k, z^k), 
    \label{eq:fsub}\\
    v^{k+1} \ = \  &  \ \underset{v \in \mathbb{R}^n_+}{\text{arg min}} \quad \mathcal{L}(f^{k+1}, v, w^k, x^k, z^k), 
    \label{eq:vsub}\\
    w^{k+1} \ = \  &  \ \underset{w \in \mathbb{R}^{n \times 2}}{\text{arg min}} \quad \mathcal{L}(f^{k+1}, v^{k+1}, w, x^k, z^k),
    \label{eq:wsub}\\
    x^{k+1} \ = \  &  \ x^k + \beta^k(Af^{k+1}-v^{k+1}),\\
    z^{k+1} \ = \  &  \ z^k + \beta^k(\mathbf{D}f^{k+1}-w^{k+1}),\\
    \beta^{k+1} \ =  \ & \sigma \beta^k, \label{eq:beta_update}
\end{align}
\end{subequations}
where $\sigma > 1$. Suggested by \cite{gu2017weighted}, the last step \eqref{eq:beta_update} is added to accelerate the numerical convergence of the ADMM algorithm. 

By assuming periodic boundary condition for the image $f$, the $f$-subproblem \eqref{eq:fsub} can be solved by fast Fourier transform \cite{wang2008new}, so it has a closed-form solution:
\begin{gather}
\begin{aligned}  \label{eq:f_update}
&{f^{k + 1}} =  \\
&{\mathcal{F}^{ - 1}}
\! 
\left( 
{\frac{
{\mathcal{F}{{(A)}^{\ast}} 
\! 
\circ
\! 
\mathcal{F}\left( {{\beta^k}{v^k} 
\! 
-
\! 
{x^k}} \right) 
\! 
- 
\! 
\mathcal{F}{{(\mathbf{D} )}^{\ast}} 
\!
\circ
\! 
\mathcal{F}
\! 
\left( {{z^k} - {\beta^k}{w^k}} \right)}
}
{{{\beta^k}\mathcal{F}(A)^{\ast} 
\! 
\circ 
\! \mathcal{F}(A) 
\! +
\!
\beta^k \mathcal{F}(\mathbf{D})^{\ast} \circ \mathcal{F}(\mathbf{D})} }}
\!
\right),
\end{aligned}
\end{gather}
where $\mathcal{F}$ is the discrete Fourier transform, $\mathcal{F}^{-1}$ is the inverse discrete Fourier transform, $\ast$ is the complex conjugate, $\circ$ is component-wise multiplication, and division is performed component-wise.

For the $v$-subproblem \eqref{eq:vsub}, the Lagrangian is separable with respect to each pixel $i$.
Thus, the $i$-th component of the solution $v^{k+1}$ satisfies the following optimality condition: 
\begin{align*}
    \frac{r+g_i}{r+v_i^{k+1}} - \frac{g_i}{v_i^{k+1}} - x_i^k - \beta^k((Af^{k+1})_i - v_i^{k+1}) = 0,
\end{align*}
or equivalently,
\begin{gather}
\begin{aligned}\label{eq:cubic}
    0 = \ & \beta^k (v_i^{k+1})^3 + (\beta^k r -\beta^k (Af^{k+1})_i - x^k_i) (v_i^{k+1})^2  \\
    & + (-\beta^k r (Af^{k+1})_i -rx^k_i +r) v_i^{k+1} - rg_i.
\end{aligned}
\end{gather}
Letting
\begin{align*}
a_0 & = -\frac{rg_i}{\beta^k},\\
a_1 & = -r (Af^{k+1})_i -\frac{rx^k_i}{\beta^k} + \frac{r}{\beta^k} ,\\
    a_2 & =  r - (Af^{k+1})_i - \frac{x^k_i}{\beta^k},
\end{align*}
the roots of the cubic equation \eqref{eq:cubic} are
\begin{align*}
    v_{i,1}^{k+1} &= -\frac{1}{3}a_2 + (S+T), \\
    v_{i,2}^{k+1} &= -\frac{1}{3}a_2 - \frac{1}{2}(S+T) + \frac{1}{2}\iota \sqrt{3}(S-T), \\
    v_{i,3}^{k+1} &= -\frac{1}{3}a_2 - \frac{1}{2}(S+T) - \frac{1}{2}\iota \sqrt{3}(S-T),
\end{align*}
where $\iota = \sqrt{-1}$ and
$$
    S = (R + D^{\frac{1}{2}})^\frac{1}{3} \quad \text{and} \quad 
    T = (R - D^{\frac{1}{2}})^\frac{1}{3}
$$
such that
$$
    R = \frac{9a_2a_1 - 27a_0-2a^3_2}{54}  \quad \text{and} \quad 
    D = \bigg ( \frac{3a_1 - a^2_2}{9} \bigg )^3 + R^2.
$$
(see \cite{neumark2014solution}).  
Since \eqref{eq:cubic} is cubic, one of the roots must be real.  Furthermore, the polynomial 
\begin{align*}
    P(\nu) = \ &\beta^k \nu^3 + (\beta^k r -\beta^k (Af^{k+1})_i - x^k_i) \nu^2  \\
    &+ (-\beta^k r (Af^{k+1})_i -rx^k_i +r) \nu - rg_i
\end{align*}
has at least one positive root because 
$P(0) < 0$ and $\displaystyle \lim_{\nu \rightarrow +\infty} P(v) = +\infty$.
As a result, the $i$-th component of the solution $v^{k+1}$ in  \eqref{eq:vsub} is given by
\begin{gather}
\begin{aligned} \label{eq:v_update}
    v_{i}^{k+1} = 
    \underset{
    v_i \in \mathcal{S}_i^{k+1}
    }
    {\text{arg min}} \ 
    &  
    (r+g_i) \log (r+v_i) - g_i \log(v_i)
    \\ 
    &
    -x_iv_i + \frac{\beta^k}{2} \left((Af)_i - v_i \right)^2
\end{aligned}
\end{gather}
where $\mathcal{S}_i^{k+1} =\{v_{i,1}^{k+1}, v_{i,2}^{k+1}, v_{i,3}^{k+1}\} \cap \mathbb{R}_+ $.

Lastly, the $w$-subproblem in \eqref{eq:wsub} is seperable with respect to each pixel $i$, so it reduces to
\begin{gather}
\begin{aligned}\label{eq:w_update}
    w_i^{k+1} = \ & \underset{w_i}{\text{arg min}} {\left\| {{w_{i}}} \right\|_1} - \alpha {\left\| {{w_{i}}} \right\|_2} \\ 
    &+ \frac{{{\beta^k}}}{2 \tau}\left\| {{w_{i}} - \left( {{{\left( {\mathbf{D} {f^{k + 1}}} \right)}_{i}} + \frac{{{{\left( {{z^k}} \right)}_{i}}}}{{{\beta^k}}}} \right)} \right\|_2^2\\
    = \ & \text{prox} \left( (\mathbf{D} f^{k+1})_{i} + \frac{z^k_{i}}{\beta^k}, \alpha, \frac{\tau}{\beta^k}  \right),
\end{aligned}
\end{gather}
where the proximal operator is defined by
\begin{align*}
    \text{prox}(x, \alpha, \beta) = \underset{y}{\text{arg min}} \|y\|_1 - \alpha \|y\|_2 + \frac{1}{2\beta} \|x-y\|^2_2.
\end{align*}
The proximal operator has a closed-form solution established in \cite[Lemma 1]{lou2018fast}.

The overall ADMM algorithm is summarized in Algorithm \ref{alg:admm}.
\begin{algorithm}[h!!!]
\caption{ADMM for 
NB recovery with AITV penalty
}
\label{alg:admm}
\begin{algorithmic}[1]
\REQUIRE Noisy, blurry image $g$, blurring operator $A$, regularization parameter $\tau$, penalty parameter $\beta^0$, penalty multiplier $\sigma > 1$.\\
    \STATE Initialize $f^0, v^0, x^0, w^0, z^0$.\\
    \STATE Set $k=0$.\\
   \WHILE{$\|f^{k}-f^{k-1}\|_2/\|f^{k}\|_2 > \epsilon$}   
   \STATE Compute $f^{k+1}$ by \eqref{eq:f_update}.
   \STATE Compute $v^{k+1}$ by \eqref{eq:v_update}.
   \STATE Compute $w^{k+1}$ by \eqref{eq:w_update}. 
   \STATE $x^{k+1} = x^k + \beta^k(Af^{k+1} - v^{k+1})$.
   \STATE  $z^{k+1} = z^k + \beta^k(\mathbf{D} f^{k+1} - w^{k+1})$.
    \STATE $\beta^{k+1}  = \sigma \beta^k$.
    \STATE $k \coloneqq k+1$.
   \ENDWHILE
   \RETURN Recovered image $\hat{f}=f^{k}$.\\
\end{algorithmic}
\end{algorithm}


\section{Experiments}
\begin{figure}[t]
\subfigure[Airplane]{\label{fig:a}\includegraphics[width=0.3\linewidth]{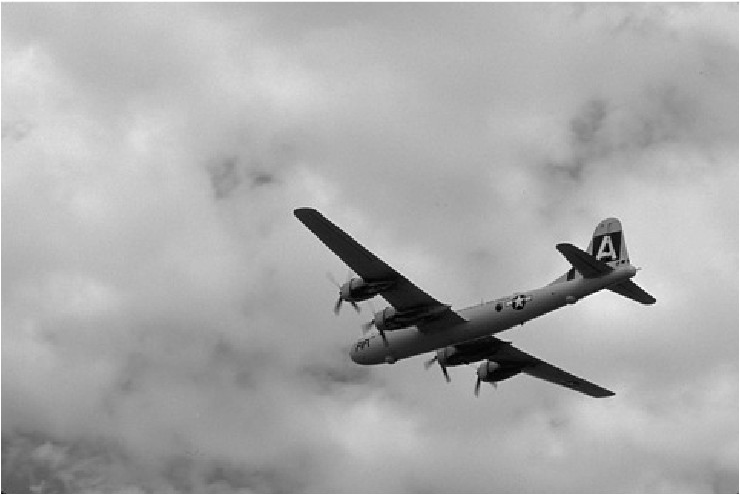}}
\subfigure[Train]{\label{fig:b}\includegraphics[width=0.3\linewidth]{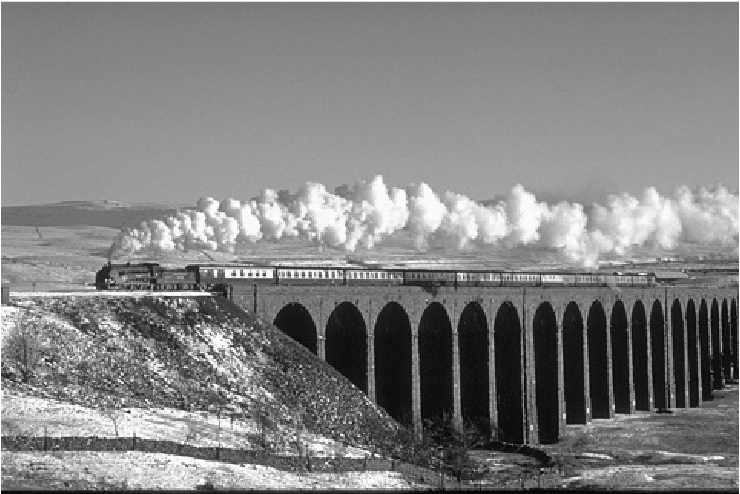}}
\subfigure[Elephant]{\label{fig:c}\includegraphics[width=0.3\linewidth]{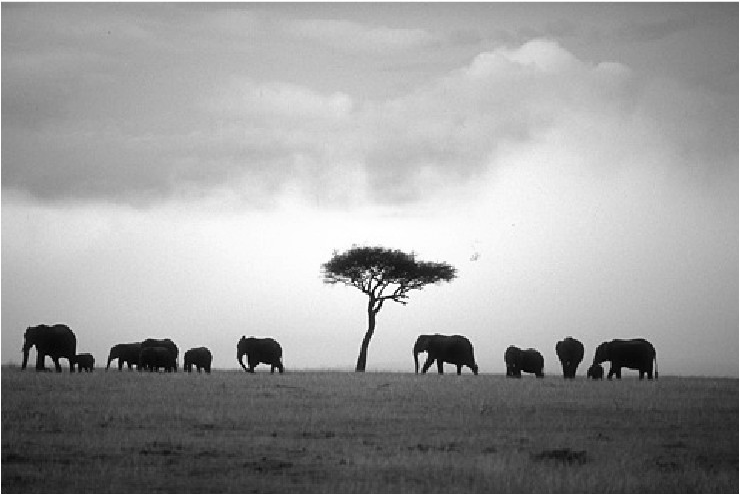}}
\subfigure[Elk]{\label{fig:d}\includegraphics[width=0.3\linewidth]{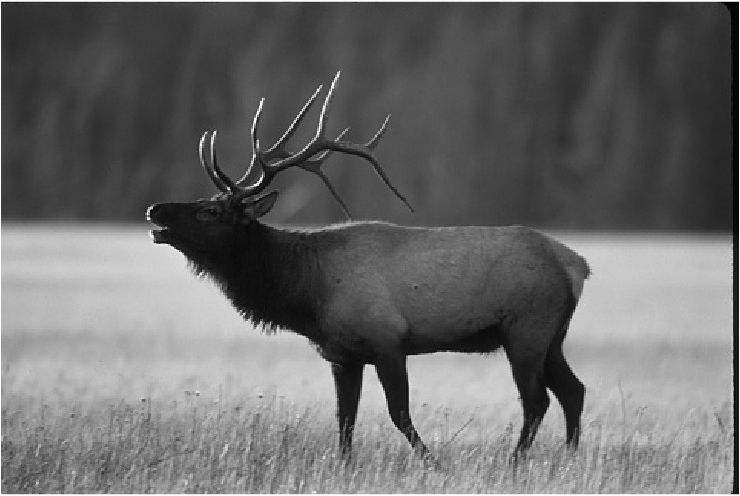}}
\hspace{0.25cm}
\subfigure[Skier]{\label{fig:e}\includegraphics[width=0.3\linewidth]{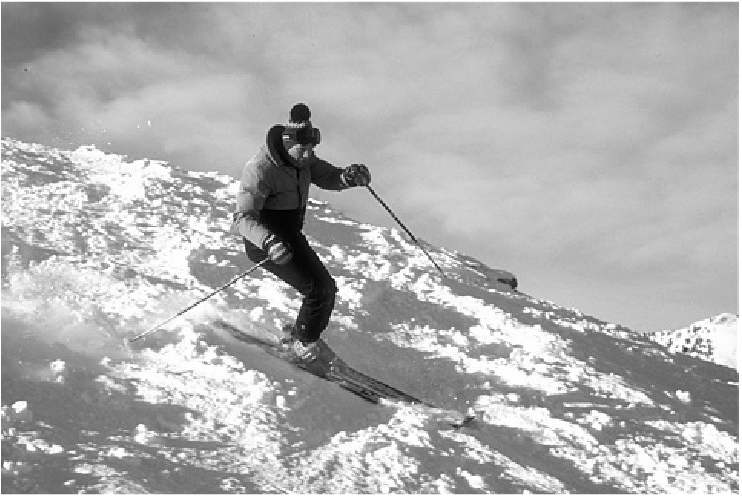}}
\caption{True images used for testing.  All five images are of pixel size $321 \times 481$. Results for (a) Airplane will be visually shown in full, while partial results for (b) Train and (c) Elephant will also be displayed. For (d) Elk and (e) Skier, only quantitative results are presented.}\label{fig:true}
\end{figure}

We conducted experiments on five different images taken from \cite{MartinFTM01} (see Fig. \ref{fig:true}), applying different NB noise levels with dispersion parameter $r \in \{ 1, 10, 25, 100, 1000\}$ to each image. Each experiment was conducted 10 times with different noisy realizations, and the average of these trials is presented.

\begin{table*}[!b]
\centering
\begin{tabular}{|l|l|
>{\raggedleft\arraybackslash}p{.85cm}|
>{\raggedleft\arraybackslash}p{.85cm}|
>{\raggedleft\arraybackslash}p{.85cm}|
>{\raggedleft\arraybackslash}p{.85cm}|
>{\raggedleft\arraybackslash}p{.85cm}|
>{\raggedleft\arraybackslash}p{.85cm}|
>{\raggedleft\arraybackslash}p{.85cm}|
>{\raggedleft\arraybackslash}p{.85cm}|
>{\raggedleft\arraybackslash}p{.85cm}|
r|
}
\cline{1-12}
\multirow{2}{*}{\ \  Image}                 
& \multirow{2}{*}{\ \ Model }
& \multicolumn{5}{c|}{PSNR} 
& \multicolumn{5}{c|}{SSIM} \\ \cline{3-12}
&
& $1$
& $10$
& $25$
& $100$
& $1000$
& $1$
& $10$
& $25$
& $100$
& $1,000$
\\ \cline{1-12}    
\multirow{2}{*}{Airplane}
& Poisson
& 13.24
& 19.25
& 24.05
& 24.11
& 24.37
& 0.40     
& 0.55     
& 0.91   
& 0.92  
& 0.92 
\\ 
\cline{2-2} 
& NB
& 23.85
& 24.05
& 24.29
& 25.02
& 25.12
& 0.84     
& 0.91     
& 0.92   
& 0.92  
& 0.93 
\\ 
\cline{1-12}
\multirow{2}{*}{Train}
& Poisson 
& 11.64
& 19.26
& 21.20
& 21.27
& 21.33
& 0.28     
& 0.57     
& 0.58   
& 0.62  
& 0.62 
\\
\cline{2-2}
& NB
& 16.03
& 20.65
& 21.66
& 22.56
& 22.61
& 0.47     
& 0.58     
& 0.61   
& 0.62  
& 0.62 
\\
\cline{1-12}
\multirow{2}{*}{Elephant}
& Poisson 
& 9.23
& 14.77
& 17.25
& 18.37
& 19.24
& 0.26     
& 0.66     
& 0.76   
& 0.81  
& 0.81 
\\
\cline{2-2}
& NB 
& 12.25
& 17.35
& 18.35
& 18.62
& 19.58
& 0.59     
& 0.78     
& 0.81   
& 0.82  
& 0.82 
\\
\cline{1-12}
\multirow{2}{*}{Elk}
& Poisson 
& 14.41
& 19.61
& 20.30
& 20.57
& 20.81
& 0.38     
& 0.71     
& 0.74   
& 0.75  
& 0.76 
\\
\cline{2-2}
& NB
& 19.33
& 20.82
& 21.45
& 22.07
& 22.21
& 0.56     
& 0.74     
& 0.75   
& 0.76  
& 0.76 
\\
\cline{1-12}
\multirow{2}{*}{Skier}
& Poisson 
& 10.09
& 19.17
& 20.60
& 21.35
& 21.71
& 0.22     
& 0.56     
& 0.60   
& 0.64  
& 0.64 
\\
\cline{2-2}
& NB
& 14.29
& 21.21
& 21.49
& 21.68
& 22.33
& 0.41     
& 0.60     
& 0.61   
& 0.64  
& 0.65
\\
\cline{1-12}
\end{tabular}
\caption{
Results for reconstruction using Poisson and negative binomial (NB) models on five different images (Airplane, Train, Elephant, Elk and Skier from the Berkeley Segmentation Dataset) at various noise levels (dispersion parameter values $r = 1, 10, 25, 100,$ and $1000$). 
Each experiment is conducted 10 times with different noisy realizations, and the average of these trials is presented. 
The first set of results (columns 3-7) correspond to the peak signal-to-noise ratio (PSNR) image quality measure, while the second (columns 8-12) correspond to the structural similarity index measure (SSIM).  
}\label{table:all}
\end{table*}

For our experiments, the linear operator $A$ is a Gaussian blur with a $10 \times 10$ window size. We compared our proposed method to the Poisson model with AITV using ADMM \cite{bui2023weighted}, which compared favorably with state-of-the-art Poisson denoising methods such as non-local PCA \cite{salmon2014poisson} and fractional-order total variation \cite{rahman2020poisson}.
Algorithm \ref{alg:admm} terminated
using the tolerance $\epsilon =  10^{-5}$. We evaluated the performance of image reconstruction using the peak signal-to-noise ratio (PSNR) and the structural similarity index (SSIM).

For the negative binomial models, several methods exist for estimating the parameter $r$, such as the method-of-moments \cite{clark1989estimation} and the maximum quasi-likelihood methods \cite{piegorsch1990maximum}. Additionally, cross-validation techniques offer a precise method for estimating the dispersion parameter \cite{gu1992cross}. Moreover, we noted that the ADMM algorithm is not sensitive to the value of $r$. That is, we may obtain very similar results when applying an approximate value of $r$. To eliminate potential biases or inaccuracies inherent in the parameter estimation process, our experiments intentionally used the exact value of the parameter $r$ as a prior.



\section{Results}

The average PSNR and SSIM metric values of the Poisson and NB models over 10 different noisy realizations at various dispersion parameter values are presented in Table \ref{table:all}.  Note that the NB model significantly outperforms the Poisson model in both metrics for low values of $r$, which correspond to settings where the observations are very noisy.  This difference decreases with increasing values of $r$, which is consistent with our understanding that the
NB distribution tends to the Poisson distribution as the dispersion parameter tends to infinity.  For $r$ values of 25, 100, and 1,000, the SSIM values for both models are comparable (with a few exceptions).  The differences in PSNR values for these $r$ values are also small but not as narrow as those for the SSIM metric.  

\begin{figure*}[t]
\centering
\begin{tabular}{ccccccc}
\hspace{-.22cm} 
\includegraphics[width=3.4cm]{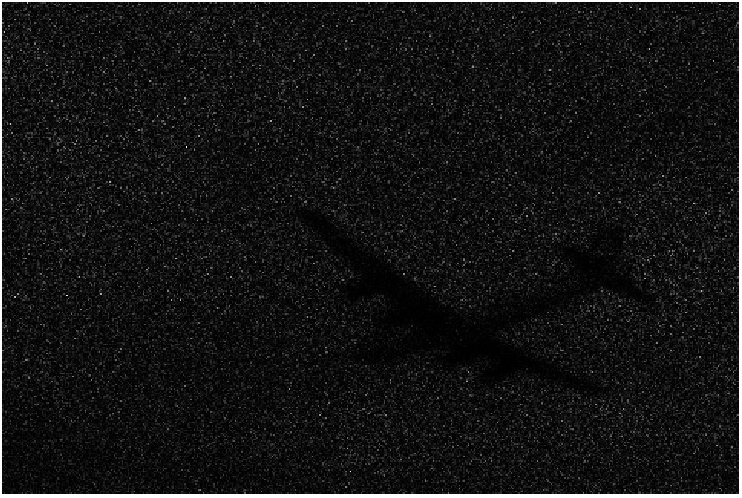} 
\hspace{-.22cm} 
&
\hspace{-.22cm} 
\includegraphics[width=3.4cm]{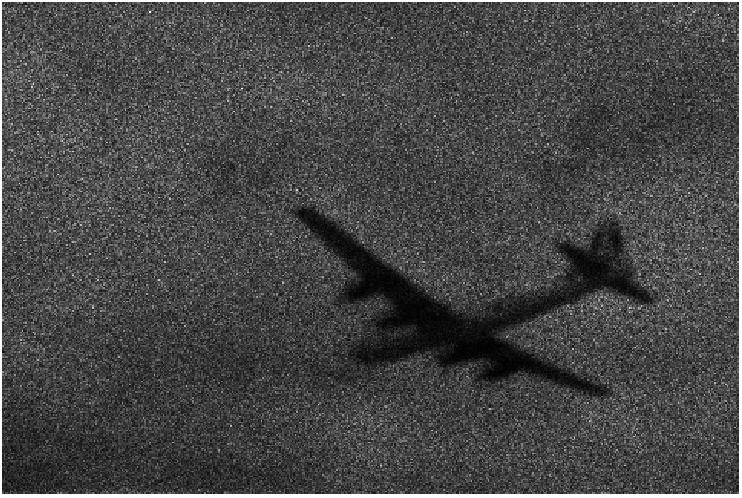} 
\hspace{-.22cm}  
&
\hspace{-.26cm} 
\includegraphics[width=3.4cm]{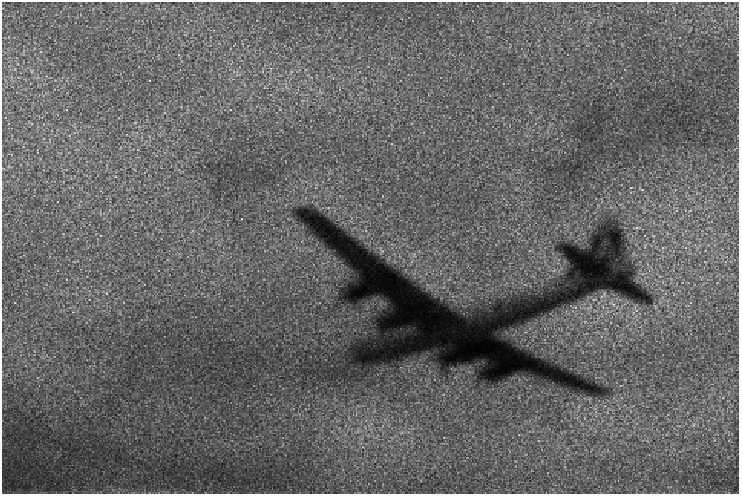} 
\hspace{-.22cm} 
&
\hspace{-.22cm} 
\includegraphics[width=3.4cm]{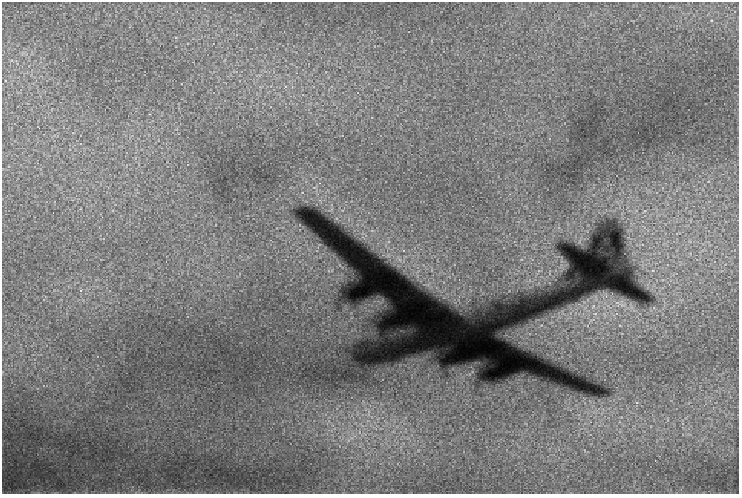} 
\hspace{-.22cm}  
&
\hspace{-.22cm} 
\includegraphics[width=3.4cm]{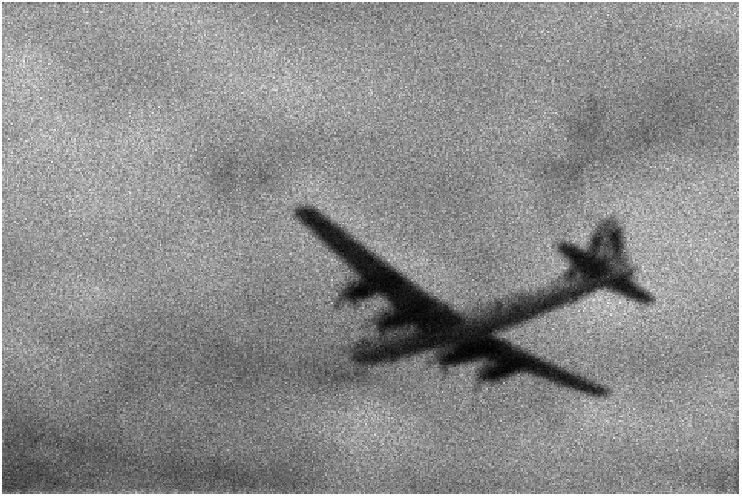} 
\hspace{-.22cm}  
\\ 
\hspace{-.22cm} 
\includegraphics[width=3.4cm]{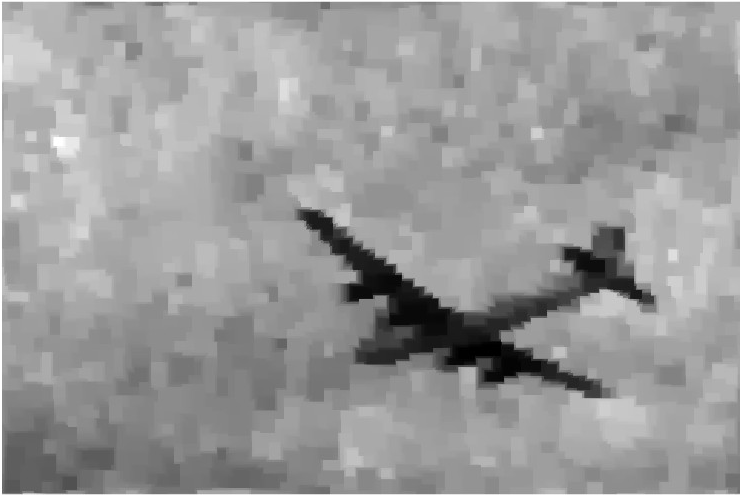} 
\hspace{-.22cm} 
&
\hspace{-.22cm} 
\includegraphics[width=3.4cm]{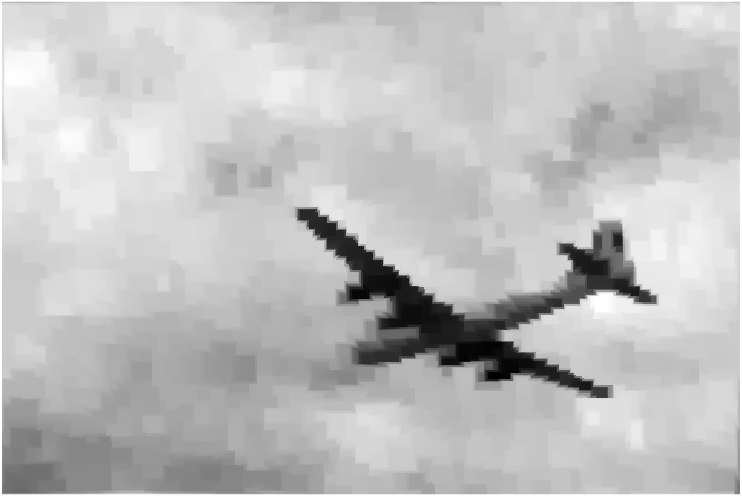} 
\hspace{-.22cm}  
&
\hspace{-.22cm} 
\includegraphics[width=3.4cm]{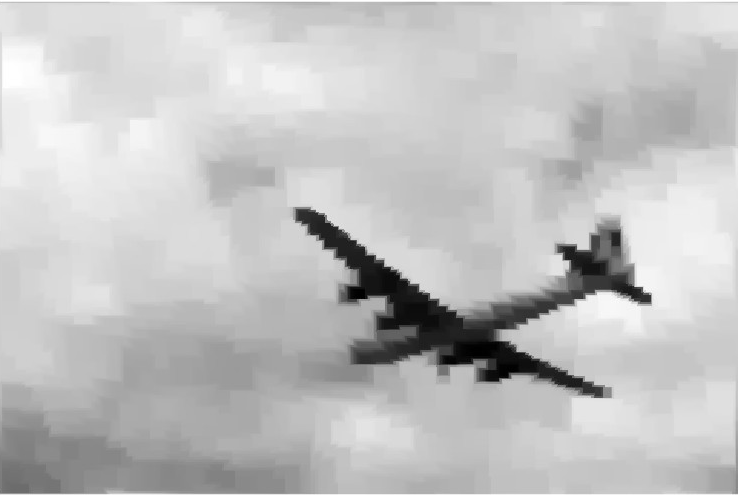} 
\hspace{-.22cm} 
&
\hspace{-.22cm} 
\includegraphics[width=3.4cm]{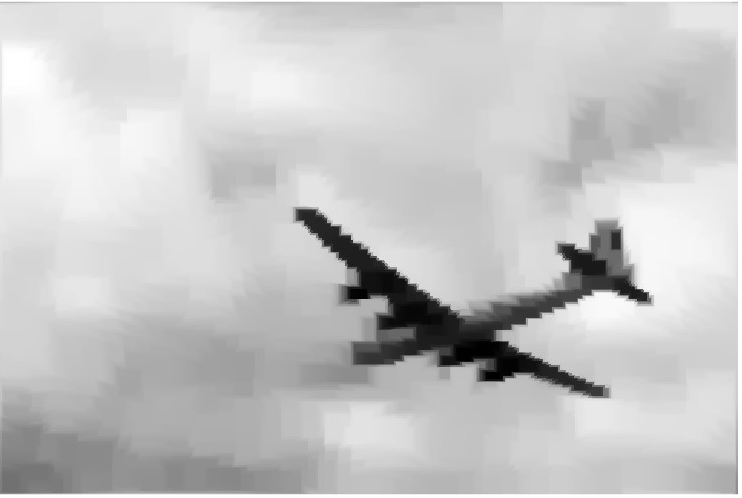} 
\hspace{-.22cm}  
&
\hspace{-.22cm} 
\includegraphics[width=3.4cm]{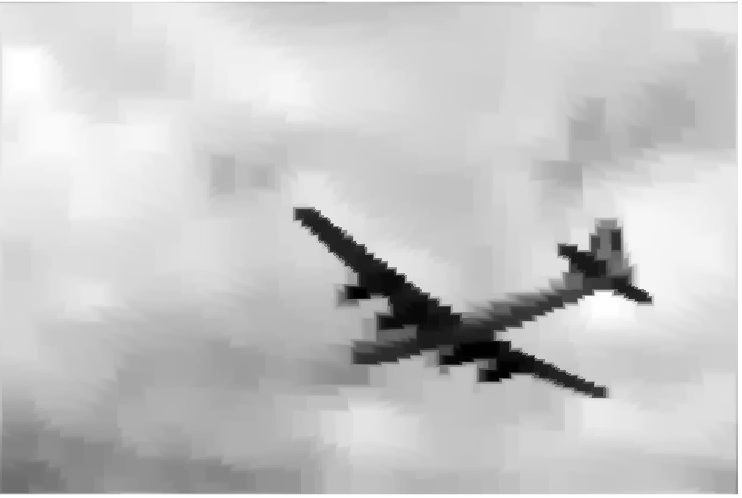} 
\hspace{-.22cm}  
\\
\hspace{-.22cm} 
\includegraphics[width=3.4cm]{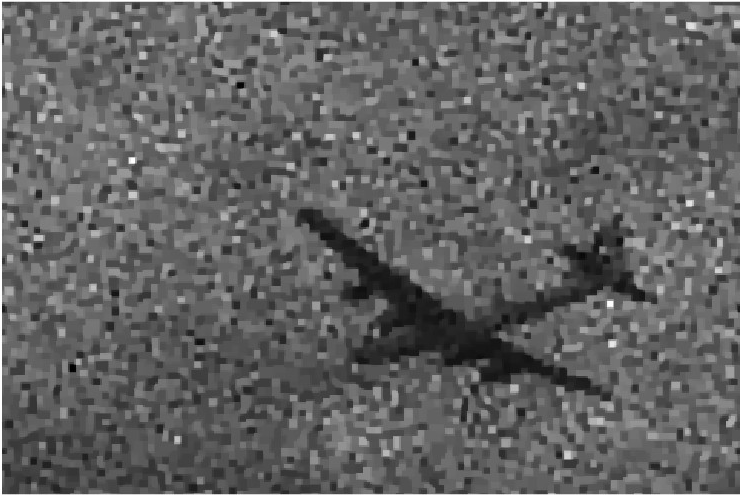} 
\hspace{-.22cm} 
&
\hspace{-.22cm} 
\includegraphics[width=3.4cm]{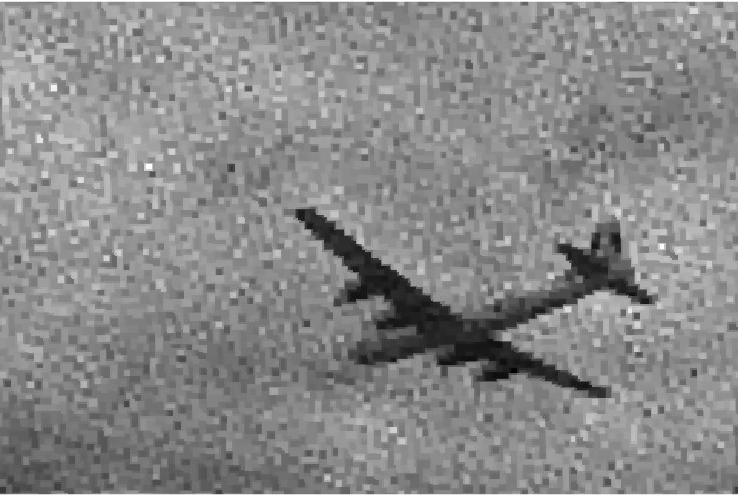} 
\hspace{-.22cm}  
&
\hspace{-.22cm} 
\includegraphics[width=3.4cm]{Images/1_r25_NB.png} 
\hspace{-.22cm} 
&
\hspace{-.22cm} 
\includegraphics[width=3.4cm]{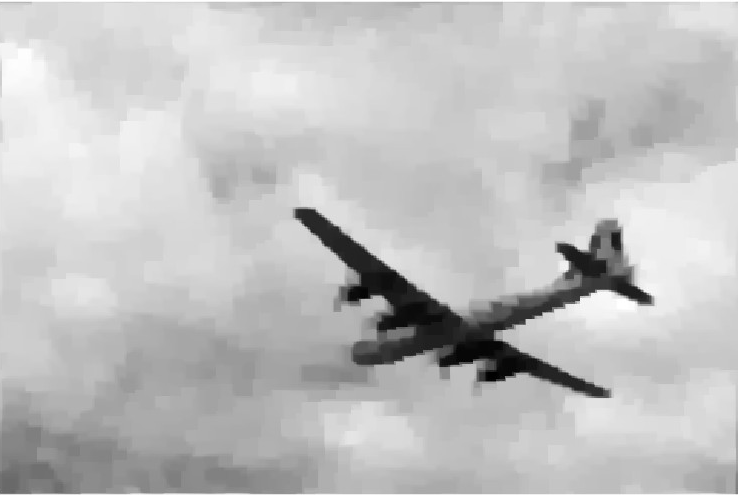} 
\hspace{-.22cm}  
&
\hspace{-.22cm} 
\includegraphics[width=3.4cm]{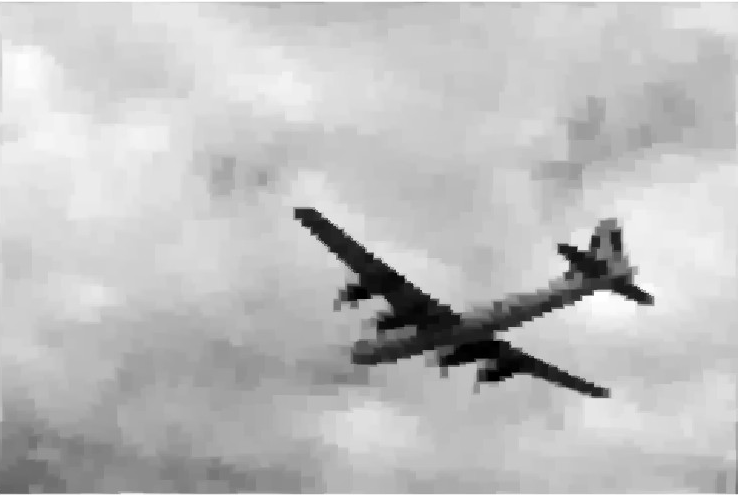} 
\hspace{-.22cm} 
\\
(a) $r=1$ & (b) $r=10$ & (c) $r=25$  
& (d) $r=100$ & (e) $r=1000$ 
\end{tabular}

\caption{Full results for the Airplane image. 
First row: noisy images.
Second row: results from the NB model.
Third row: results from the Poisson model. Each column corresponds to a different noise level. 
}
\label{fig:airplane}
\end{figure*}

We present representative reconstructions for the Airplane image (see Fig.\ \ref{fig:true}(a)) for all five dispersion parameter values $r$ in Fig.\ \ref{fig:airplane}.  The first row correspond to the noisy blurry images; the second, to the NB model reconstructions; and the third, to the Poisson model reconstructions.  Note how low the pixel intensities are for observations with low dispersion parameter values ($r = 1$ and $r = 10$).  Regarding reconstructions, note how speckled the skies are in the Poisson reconstructions at these low $r$ value settings while the reconstruction of the clouds are smoother in the NB reconstruction. 

\begin{figure}[h!]
    \centering
    \begin{tabular}{cc}
    \hspace{-.2cm}
    \includegraphics[width=0.23\textwidth]{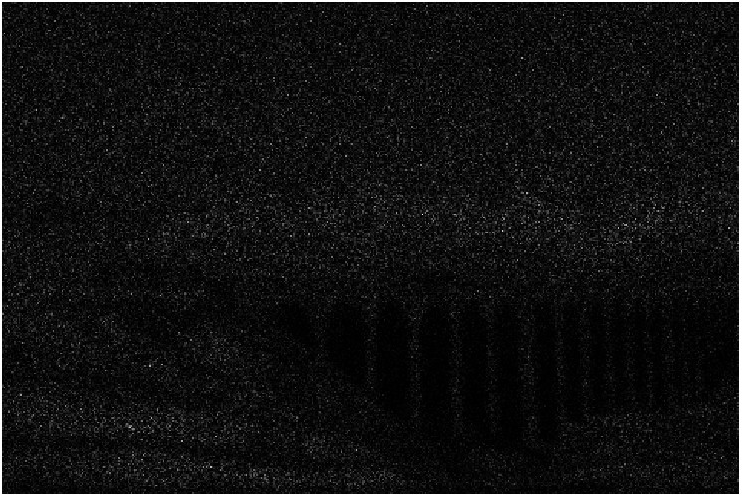}
    \hspace{-.2cm} &  \hspace{-.2cm}   \includegraphics[width=0.23\textwidth]{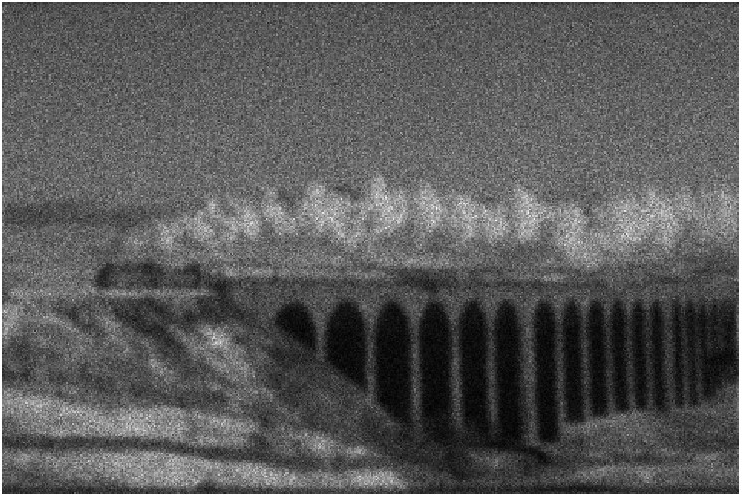}
    \\ 
    \hspace{-.2cm}
    \includegraphics[width=0.23\textwidth]{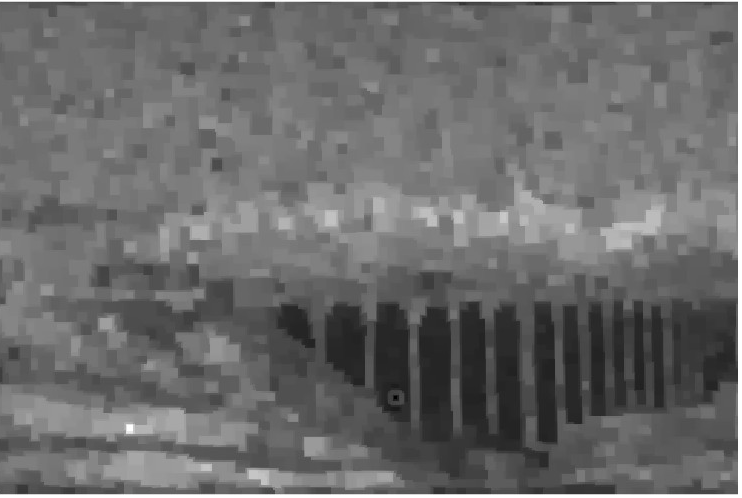}
    \hspace{-.2cm} &  \hspace{-.2cm}    \includegraphics[width=0.23\textwidth]{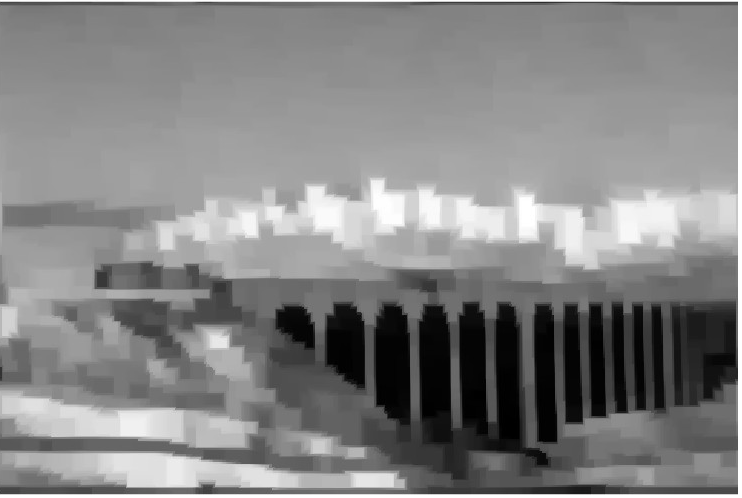}
    \\ 
    \hspace{-.2cm}
    \includegraphics[width=0.23\textwidth]{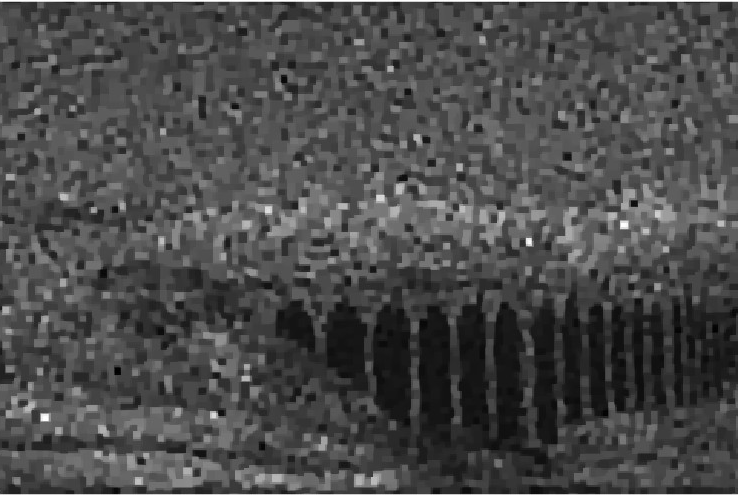}
    \hspace{-.2cm} &  \hspace{-.2cm} \includegraphics[width=0.23\textwidth]{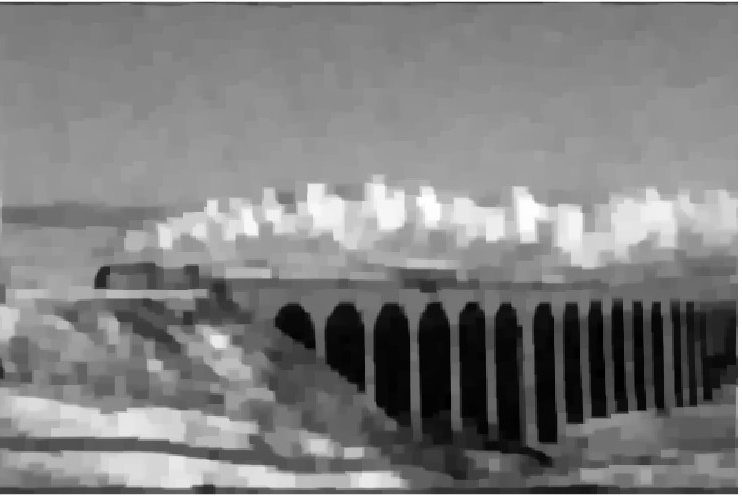} \\
    $r = 1$ & $r = 100$ 
    \end{tabular}
    \caption{Partial results of the Train image. First row: noisy images.  Second row: results from the NB model. Third row: results from the Poisson model. The first column correspond to $ r = 1$ and the second column correspond to $r = 100$.}
    \label{fig:train}
\end{figure}
\begin{figure}[h!]
    \centering
    \begin{tabular}{cc}
    \hspace{-.2cm}
    \includegraphics[width=0.23\textwidth]{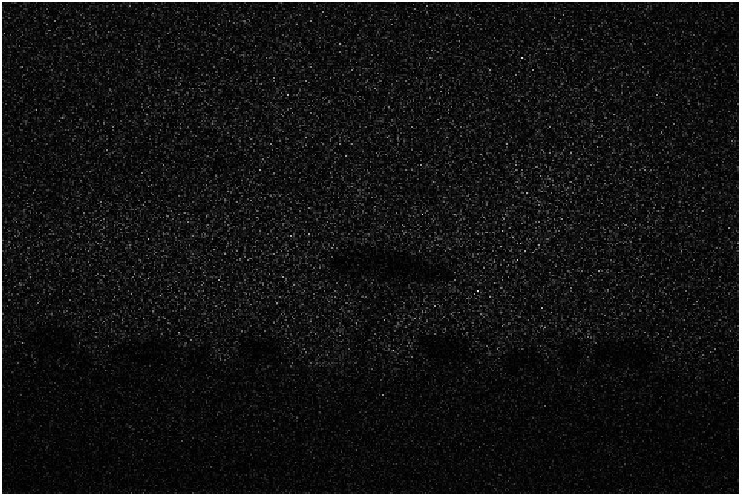}
    \hspace{-.2cm} &  \hspace{-.2cm}   \includegraphics[width=0.23\textwidth]{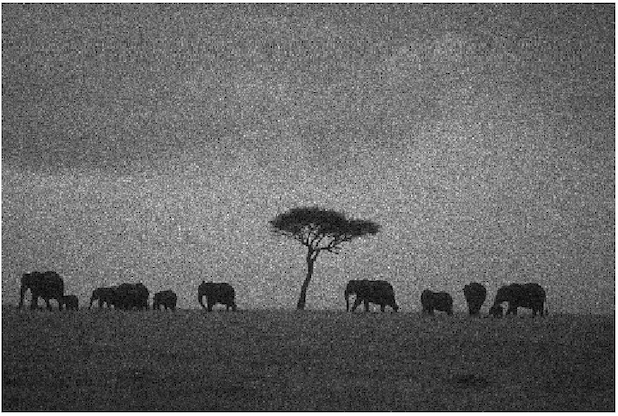}
    \\ 
    \hspace{-.2cm}
    \includegraphics[width=0.23\textwidth]{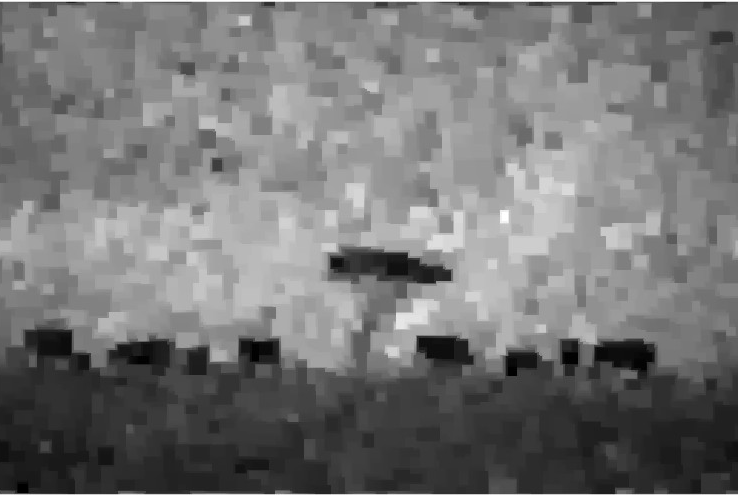}
    \hspace{-.2cm} &  \hspace{-.2cm}    \includegraphics[width=0.23\textwidth]{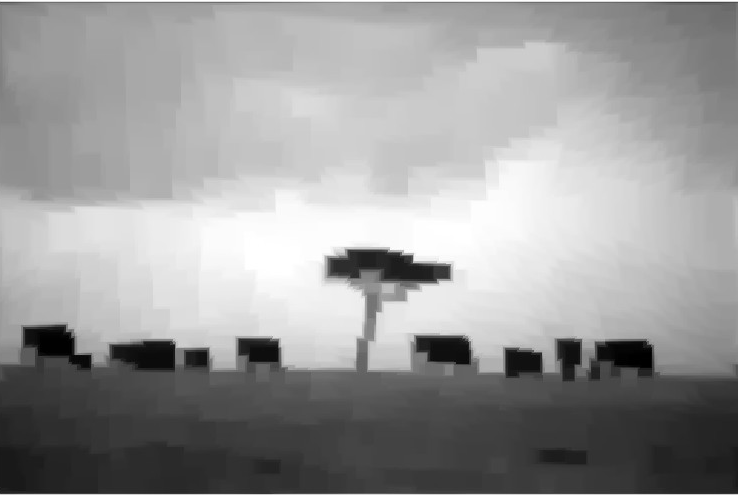}
    \\ 
    \hspace{-.2cm}
    \includegraphics[width=0.23\textwidth]{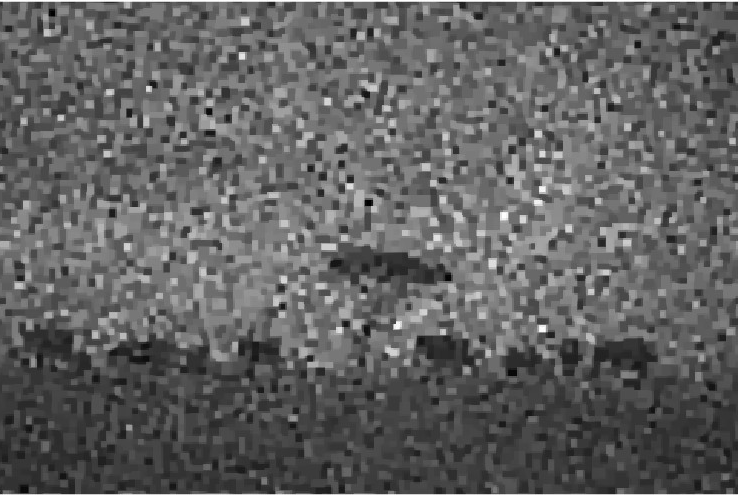}
    \hspace{-.2cm} &  \hspace{-.2cm} \includegraphics[width=0.23\textwidth]{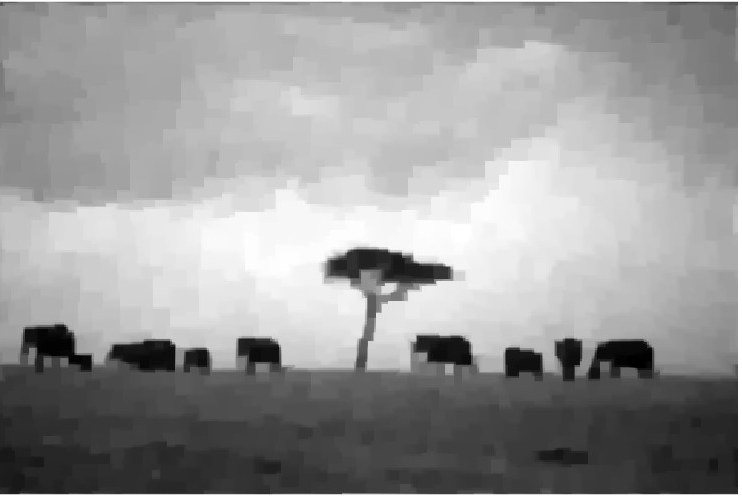} \\
    $r = 1$ & $r = 100$
    \end{tabular}
    \caption{Partial results of the Elephant image. First row: noisy images.  Second row: results from the NB model. Third row: results from the Poisson model. The first column correspond to $ r = 1$ and the second column correspond to $r = 100$.}
    \label{fig:elephant}
\end{figure}

We also present representative reconstructions for the the Train and Elephant images (see Figs.\ \ref{fig:true}(b) and \ref{fig:true}(c)) in Fig.\ \ref{fig:train} and Fig.\ \ref{fig:elephant}, respectively, for $r = 1$ and $r = 100$.  Note the low pixel intensities for the $r =1$ setting.  Similar to the reconstructions for the Airplane image, the Poisson reconstructions are speckled while the NB reconstructions are smoother.  Note how similar the reconstructions are for the higher dispersion parameter value $r = 100$.


\section{Conclusion}
We proposed an NB model with AITV regularization to recover an image that has overdispersed Poisson noise. To solve this model, we developed an efficient ADMM algorithm whose subproblems have closed-form solutions. Since the NB model is a generalization of the Poisson model, we attained better results than the Poisson model for images corrupted by overdispersed Poisson noise. Our experiments demonstrated that the NB model return higher PSNRs and SSIMs than the Poisson model at various levels of NB noise. 
As expected,  
the results for the NB and Poisson models become more similar
as the dispersion parameter increases
since the Poisson distribution is a limiting case of the NB distribution.


\bibliographystyle{IEEEtranS}
\bibliography{ref}

\end{document}